%% file: main.tex
\documentclass[journal]{vgtc}                     

\onlineid{0}

\vgtccategory{Research}

\vgtcpapertype{please specify}

\title{Direct vs. Score-based Selection: Understanding the Heisenberg
Effect in Target Acquisition Across Input Modalities in Virtual Reality}

\author{%
  \authororcid{Linjie Qiu}{0009-0008-9204-5690},
  \authororcid{Duotun Wang}{0009-0005-4393-5230}, 
  \authororcid{Boyu Li}{0009-0006-7265-9236},
  \authororcid{Jiawei Li}{0009-0000-6593-2958},
  \authororcid{Yulin Shen}{0000-0003-1635-6661},
  \authororcid{Zeyu Wang}{0000-0001-5374-6330}, and 
  \authororcid{Mingming Fan}{0000-0002-0356-4712}
}

\authorfooter{
  \item
    Linjie Qiu, Duotun Wang, Jiawei Li, Yulin Shen, Zeyu Wang, Mingming Fan are with The Hong Kong University of
    Science and Technology (Guangzhou). E-mail: lqiu250, dwang866, jli526, yshen654@connect.hkust-gz.edu.cn
  \item
  	Boyu Li, Zeyu Wang, Mingming Fan are with The Hong Kong University of
    Science and Technology. E-mail: blibr, zeyuwang, mingmingfan@ust.hk. \\
  \item Mingming Fan is the corresponding author.
}

\abstract{%
  Target selection is a fundamental interaction in virtual reality (VR). But the act of confirming a selection, such as a button press or pinch, can disturb the tracked pose and shift the intended target, which is referred to as the Heisenberg Effect. Prior research has mainly investigated controller input. However, it remains unclear how the effect manifests in the bare-hand input and how score-based techniques may mitigate the effect in different spatial variations. To fill the gap, we conduct a within-subject study to examine the Heisenberg Effect across two input modalities (i.e., controller and hand) and two selection mechanisms (i.e., direct and score-based). Our results show that hand input is more susceptible to the Heisenberg Effect, with direct selection more influenced by target width and score-based selection more sensitive to target density. Based on previous vote-oriented technique and our temporal analysis, we introduce weighted VOTE, a history-based intention accuracy model for target voting, that reweights recent interaction intent to counteract input disturbances. Our evaluation shows the method improves selection accuracy compared to baseline techniques. Finally, we discuss future directions for adaptive selection methods.
}


\teaser{
  \centering
  \includegraphics[width=1.0\linewidth]{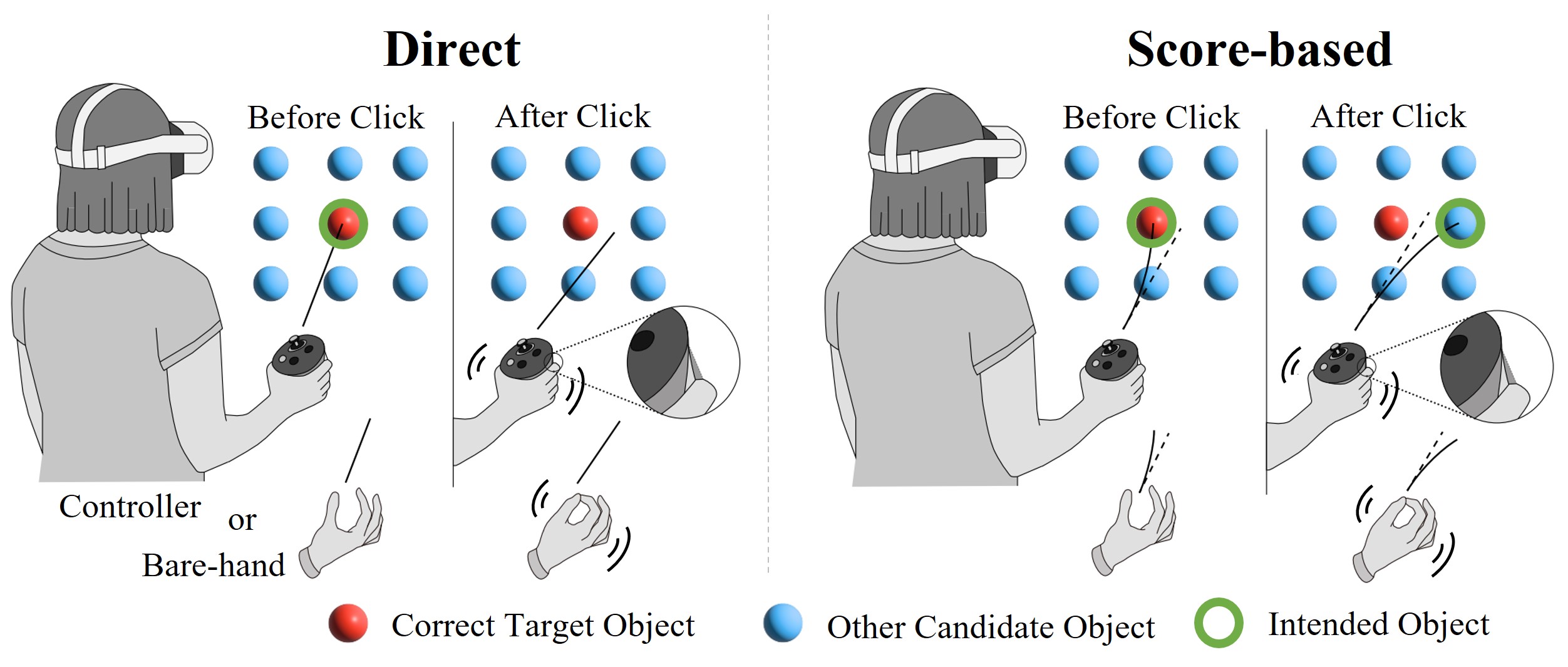}
  \caption{Both direct and score-based selection are affected by action disruption during confirmation. While score-based selection infers user intent to reduce the error rate, it nearly always snaps to a certain object, making errors often irreversible compared to direct selection. Our user study (\textit{N} = 24) examines how direct and score-based selection methods with different input modalities (i.e., controller and hand inputs) are influenced by the Heisenberg Effect on task performance and user experience.}
  \label{fig:teaser}
}

\graphicspath{{figs/}{figures/}{pictures/}{images/}{./}} 

\usepackage{tabu}                      
\usepackage{booktabs}                  
\usepackage{lipsum}                    
\usepackage{mwe}                       
\usepackage{times}\usepackage{amsmath} 
\usepackage{mathptmx}                  
\usepackage{makecell}
\usepackage[normalem]{ulem}

\begin{document}



\input{1-Introduction}

\input{2-Relatedwork}
\input{3-Experiment}

\input{4-Result}

\input{5-MitigationTechniques}
\input{6-Discussion}
\input{7-Conclusion}

\acknowledgments{%
This research is partially funded by 2025 Guangdong Undergraduate University Teaching Quality and Teaching Reform Project, AI Research and Learning Base of Urban Culture under Project 2023WZJD008, Guangdong Provincial Key Lab of Integrated Communication, Sensing and Computation for Ubiquitous Internet of Things ( No. 2023B1212010007), and the Project of DEGP (No. 2023KCXTD042). This work is also partially supported by the National Natural Science Foundation of China (No. U25A20384). And we thank the support from Red Bird MPhil Program at HKUST(GZ).
}


\bibliographystyle{abbrv-doi-hyperref}

\bibliography{ref}

\end{document}

%% file: 1-Introduction.tex
\firstsection{Introduction}
\maketitle


Target selection, as a fundamental interaction task in virtual reality (VR) and augmented reality (AR), underpins a wide range of applications such as object manipulation and task execution. In current commercial VR systems, users perform selection tasks using various input modalities~\cite{Quest3,Pico4}, with handheld controllers and bare-hand tracking being the most prevalent due to their intuitiveness and versatility~\cite{HandorController:2023:TVCG}. However, both input modalities involve the tracking of discrete actions like button presses or finger pinches, which can disturb the user's acquisition of the target and introduce unintended spatial offsets~\cite{Heisenbergv1:2001:HCII,HeisenbergEffect:CHI:2020}. This phenomenon, known as the \textbf{Heisenberg Effect} in spatial computing~\cite{HeisenbergEffect:CHI:2020}, describes how the discrete action of selection interferes with the target being observed or targeted, thereby affecting selection performance.

While the Heisenberg Effect has been systematically evaluated in direct controller input, several studies have informally noted that this effect may be more pronounced during hand-based interactions, likely due to greater movement variability and the lack of physical constraints~\cite{RedirectRay:2024:SUI,RestfulRaycast:2025:DIS,FocalSelect:2024:TVCG}. However, the impact of this effect on bare-hand input has yet to be formally and quantitatively evaluated. Beyond input modalities, recent works have adopted score-based selection, such as snap-to raycasting~\cite{IntenSelect:2005:EGEV,IntenSelectPlus:2024:TVCG}, focal selection~\cite{FocalPoint:IMWUT:2023,FocalSelect:2024:TVCG} and bubble-based mechanism~\cite{BubbleRay:2020:VR,GazeBubble:2025:IJHCI}, to mitigate the Heisenberg Effect. These score-based selection methods estimate user intent dynamically, reducing the need for precise alignment and avoiding the risk of selecting empty targets~\cite{IntenSelect:2005:EGEV}, reducing the extent to which target width affects performance as predicted by Fitts’s law~\cite{FittsLaw:1954:JEP}. Unlike direct selections, score-based approaches greatly reduce the likelihood of void selections. As a result, when errors occur, they are more likely to cause irreversible or destructive actions immediately. Despite the widespread use of such methods, the influence of the Heisenberg Effect on the performance of score-based selections in VR remains under-examined, and there is a paucity of research focusing on the dense and occluded settings~\cite{MINIMAP:2023:VR,GeneralModel:2006:3DUI,FullyOccluded:2020:TVCG}. This gap motivates us to investigate how and to what extent the Heisenberg effect varies across different input modalities and mechanisms.

In this work, we first extend the analysis of the Heisenberg effect in target selection to both controller and hand input with direct and score-based mechanisms. It is worth noting that the pinch, the most commonly used gesture to select distant objects~\cite{PinchLens:2023:ISMAR,OculusDistance:2023}, is adopted as the confirmation action in our study. We find that during ballistic selections, the Heisenberg effect accounts for 80.57\% of selection errors for direct controller selection, 47.50\% for score-based controller selection, 71.86\% for direct hand selection, and 46.86\% for score-based hand selection. From the perspective of the Heisenberg Effect, our results offer a novel explanation for why the selection accuracy of hand tracking is inferior to that of controller tracking. We also observe that direct selection is more sensitive to target width, while score-based mechanisms can partially compensate for the Heisenberg effect. However, their effectiveness varies significantly depending on target density (described as \textit{target spacing} in our experiment). Furthermore, we find that as task difficulty increases, the Heisenberg Magnitude decreases, which may be attributed to users subconsciously stabilizing their hands in response to perceived difficulty, which reduces unintended movements.

To track intention shifts and accuracy trends across different selection techniques over time, we fit a third-degree polynomial to model the probability of correctly matching the intended object to the target (Section~\ref{temporalanalysis}). Our evaluations indicate that although both direct and score-based hand input exhibit lower final selection accuracy, the temporal records reveal intermittent moments of high precision. Building on the insights and classic history-based selection techniques such as VOTE~\cite{VOTE:2018:IJHCS} and BackTracer~\cite{BackTracer:2023:IJHCS}, our quantitative analysis suggests that historical data should not contribute equally to final selection decisions. Accordingly, we refine the fitted temporal accuracy function into VOTE by incorporating a weighted function trained on the collected data. Our ablation study reveals that all input methods benefit from Weighted VOTE, with the most pronounced improvement observed in score-based hand input. The selection error rate is reduced substantially, from 21.96\% to 7.54\%, reaching a level comparable to controller input, where the score-based controller input achieves an 8.18\% error rate after improvement. We further analyze individual intention accuracy histories to guide personalized adjustments. Although no significant improvement is observed, we outline implications for future adaptive methods in VR. This study offers the following contributions:
\begin{itemize}
    \item We extend the systematic analysis of the Heisenberg Effect in VR to different input modalities and selection mechanisms.
    \item We analyze the temporal distribution of users’ intention shift and accuracy variation in selection events.
    \item We introduce and evaluate a weighted VOTE method to mitigate the Heisenberg effect by leveraging historical intention.
\end{itemize}

%% file: 2-Relatedwork.tex
\section{Related Work}
\label{relatedWork}

To develop a general understanding of the Heisenberg Effect, we first review prior work that attempts to mitigate the effect and its implications across input modalities (Section~\ref{relatedworks: Heisenberg}). We then emphasize interaction techniques that leverage score-based heuristics and users’ intention history for the selection task, which demonstrated to be especially efficient and directly inform our approach (Section~\ref{relatedworks: history}).

\subsection{Heisenberg Effect in Spatial Interaction}
\label{relatedworks: Heisenberg}
The Heisenberg effect, first noted by Bowman et al.~\cite{Heisenbergv1:2001:HCII} as a side effect of manipulating spatially tracked input devices (STIDs), was further analyzed in depth by Wolf et al.~\cite{HeisenbergEffect:CHI:2020} for controller tracking in the target selection task. Through a systematic analysis of user behaviors, they identified key contributing factors that affect this effect (i.e., target width), and defined the Heisenberg Error as \textit{``starting a selection inside the target but ending outside due to a disturbance of the input device.''} Their investigations also outlined that discrete inputs, such as a button press, can physically disturb the STID, shifting the selection point during ray-cast interaction. While the Heisenberg effect is a well-explored concern for direct controller selection~\cite{BackTracer:2023:IJHCS,FocalSelect:2024:TVCG,SelectionSurvey:2024:CUSR}, its impact on bare-hand tracking remains largely unevaluated and bare-hand tracking is also prone to disturbances from discrete actions like pinch~\cite{RedirectRay:2024:SUI,RestfulRaycast:2025:DIS}.

Furthermore, from the perspective of input modality, controller and bare-hand are the two most widely used for object selection and manipulation in VR~\cite{TradeOffinVR:2022:MTI}. Controllers typically feature joysticks and buttons~\cite{ControllerSurvey:2019:CGF}, offering benefits such as physical haptic feedback, ergonomics, and precise, consistent control~\cite{HandorController:2023:TVCG}. Meanwhile, hand tracking often relies on vision-based algorithms~\cite{GestureReview:2020:JOI} to detect gestures, enabling natural, device-free interaction that enhances users’ sense of presence and body ownership~\cite{EFree:2020:MTI,EffectHandSize:2019:VR}. However, despite its intuitiveness, bare-hand input consistently underperforms in target selections compared to controllers in terms of selection efficiency and accuracy~\cite{EFree:2020:MTI,ComparisonConHand:2023:AS,FocalSelect:2024:TVCG}, especially for distant object interactions~\cite{GOGO:UIST:96, NinjaHands:CHI:2021}. Prior studies primarily attribute this inferiority to the limited reliability and precision of tracking algorithms~\cite{Mid-AirPointing:2018:CHI, SurveyVRSelect:2013:CG}. In this work, we investigate the role of the Heisenberg effect as an overlooked factor for performance degradation. We conduct a quantitative analysis across input modalities to determine the extent to which it compromise selection performance and user experience.

\subsection{Score-based Selections in VR}
\label{relatedworks: history}

Besides techniques based on classic ray-casting, score-based selection techniques have been developed to improve target acquisitions in VR, particularly for interacting with distant or dynamic objects~\cite{IntenSelectPlus:2024:TVCG, GazeBubble:2025:IJHCI, OculusRayInteactions:2025}. These methods assign an intention score for each potential target and snap to the user's desired one. Early approaches like IntenSelect~\cite{IntenSelect:2005:EGEV,IntenSelectPlus:2024:TVCG} and StickyRay~\cite{SticyRay:CVG:2006} compute such a score based on the angle between the user's ray direction and targets. This principle has been incorporated into practical development workflows (e.g., Meta's \textit{Distance Hand Grab Interactor}~\cite{OculusDistance:2023}). A key extension of this concept is focal-based selections~\cite{FocalPoint:IMWUT:2023,FocalSelect:2024:TVCG}, which integrates spatial distance into the intent model and adapts it across various input modalities. However, a significant limitation of these score-based techniques in dense environments has been noted by FocalSelect~\cite{FocalSelect:2024:TVCG} and BubbleRay~\cite{BubbleRay:2020:VR}. Our experimental results corroborate this and the Heisenberg Effect may inadvertently destabilize selections, making it harder to acquire the intended target.

Beyond snap-to methods, history-based scoring has emerged as another type of score-based technique to infer users' intent. For example, VOTE~\cite{VOTE:2018:IJHCS} selects the object with the highest occurrence in a fixed-time period. BackTracer~\cite{BackTracer:2023:IJHCS} enhances it by dynamically adjusting the period duration based on the perceived stability of targets. To mitigate the Heisenberg Effect, Wolf et al.~\cite{HeisenbergEffect:CHI:2020} have leveraged historical interaction traces to infer the originally intended target, compensating for spatial or temporal disruptions. However, this method appears to increase many false positive selection errors, making it not robust in many cases.

Our work builds upon these efforts by incorporating weighted historical temporal inputs. Through quantitative analysis in Section~\ref{experiment}, we fitted functions to model the temporal contributions of historical inputs. These functions serve as a mitigation strategy against the Heisenberg Effect and are evaluated through an ablation study, demonstrating the effectiveness across various input modalities and selection mechanisms.

%% file: 3-Experiment.tex
\section{Experiment}
\label{experiment}
In this section, we investigate and analyze how and to what extent various input modalities and selection mechanisms are influenced by the Heisenberg effect during target selection tasks. We further analyzed pre-selection user behavior and intention history. 

\subsection{Participants}

We recruited 24 participants (11 females and 13 males), with an average age of 25.0 (SD = 2.74). All individuals self-identified as right-handed and exhibited unimpaired vision and color discrimination abilities. 15 of them had experience within VR as designers or developers, while others have less or no experience with VR devices. Each participant received a compensation of 50 RMB for participation. This study was approved by the Institutional Ethical Review Board (IRB), and informed consent was obtained from participants.

\subsection{Apparatus}

The experiments were developed in Unity 2022 LTS, ran on Quest 3, and driven by a Windows 10 desktop (CPU: AMD Ryzen 7 5800H, GPU: Nvidia GeForce RTX 3060). Bare-hand and controller tracking were supported with the Meta Oculus SDK.

To ensure the fairness of our experiment, we implemented a unified white ray as the visual indicator for direct ray-casting tracking, which was supported for both hand and controller inputs by the Oculus SDK's raycasting interaction framework~\cite{OculusRayInteactions:2025}. Following the pointing task in Wolf et al.~\cite{HeisenbergEffect:CHI:2020}, users performed target selection via the trigger button on the Quest 3 controller. The Oculus SDK registers the trigger's pressure on a continuous scale from 0 (not pressed) to 1 (fully pressed), with selection events triggered upon a full press. For bare-hand input, confirmation was triggered via thumb–index pinch gestures, detected by the built-in pose recognition system~\cite{OculusPoseDetection:2023:Oculus}. Details on detecting the start of a pinch action are provided in Section~\ref{PinchAction}.

We implemented Intenselect~\cite{IntenSelect:2005:EGEV}, a classical snap-to selection technique in VR, to support the score-based selection in two modalities. All score-based methods used a unified snap-to ray as the visual indicator and followed a spatiotemporal score calculation, with the top-ranked object~\cite{IntenSelect:2005:EGEV} selected. The scoring function is defined as follows: 
\begin{equation}
    s_t = s_{t-1} \gamma + \left(1 - \frac{\eta(t)}{\varepsilon} \right)(1 - \gamma),
\end{equation}
where $s_t$ denotes the score at time $t$, $\eta(t)$ is the angle between the original raycasting direction and the direction from the ray origin to the target’s center at time $t$, $\epsilon$ is the angular threshold, and $\gamma$ is the temporal decay factor, fixed at 0.5.

\begin{figure}
    \centering
    \includegraphics[width=1.0\linewidth]{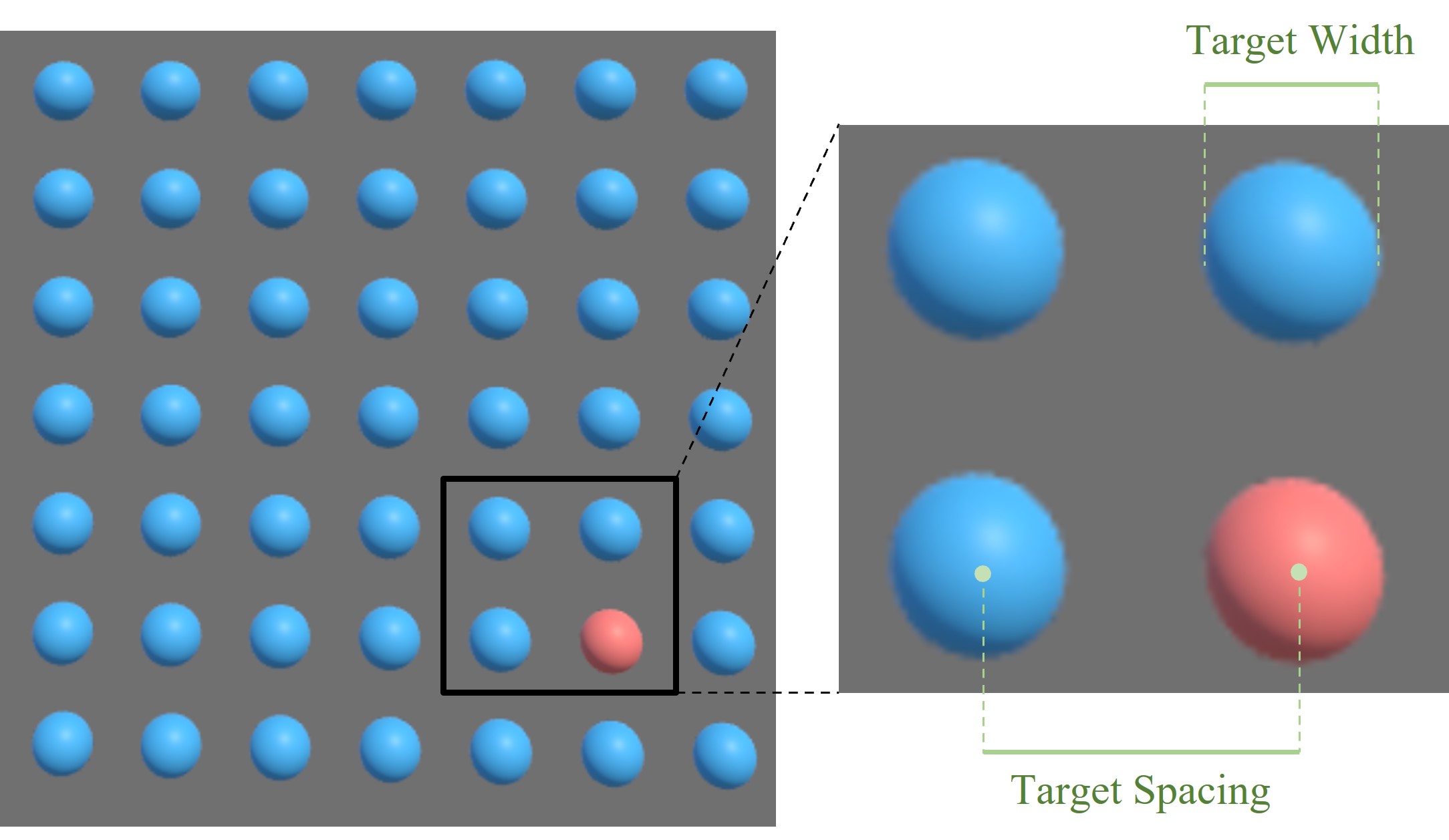}
    \caption{An example from our experimental design in Unity, which includes nine combinations of target width and spacing. In this example, the target width is 42 cm, and the target spacing is 70 cm. The distance between the user and targets is fixed at 8 meters.}
    \label{fig:setup}
\end{figure}
\subsection{Testing Variables}

\subsubsection{Independent Variables}
Attributing the Heisenberg effect to physical interaction, we hypothesized that the input method significantly influences the selection performance and its magnitude \textbf{(H1)}. As shown in Figure~\ref{fig:teaser}, we evaluated four combinations of input modality and selection mechanism as \textbf{Input Techniques} to analyze the Heisenberg Effect: \textit{(a) direct selection with controller input (DC)}, \textit{(b) direct selection with hand input (DH)}, \textit{(c) score-based selection with controller input (SC)}, and \textit{(d) score-based selection with hand input (SH)}. Following the experimental setup in Wolf et al.~\cite{HeisenbergEffect:CHI:2020} and informal findings in BubbleRay~\cite{BubbleRay:2020:VR}, we hypothesized that the target size and environmental density can be significant variables for the selection performance and magnitude of the Heisenberg Effect \textbf{(H2)}. Thus, we introduced two environmental variables: \textbf{Target Width} \textit{(14, 28, 42 cm)} and \textbf{Target Spacing} between the targets \textit{(30, 50, 70 cm)}. The user's distance from the selection targets is fixed at 8 meters, enabling the measurement of the spatial shift of ray-casting using an offset angle relative to the participant. We show the example experiment setup in Figure~\ref{fig:setup}.

\subsubsection{Dependent Variables}
\label{PinchAction}
We collected the following dependent variables: \textbf{Selection Time}, \textbf{Overall Error}, \textbf{Heisenberg Error}, and \textbf{Heisenberg Magnitude}.

\begin{itemize}
    \item The \textbf{Selection Time} was defined as the time taken to complete a single selection action.
    \item The \textbf{Overall Error} was computed as the percentage of total selections that resulted in missed or incorrect targets.
    \item The \textbf{Heisenberg Error} captured the proportion of selections where the selection began inside the target area but concluded with a miss or incorrect selection.
    \item The \textbf{Heisenberg Magnitude} was measured as the angular offset between the start and end of the selection action.
\end{itemize}

To accurately detect the start of a pinch gesture, we used two key signals: the rate of change in distance between the thumb and index finger (\textit{velocityTI}) and the palm's rotational velocity (\textit{velocityRot}). The detection process involves two steps:
\begin{itemize}
    \item \textbf{Identify the potential pinch onset:}  We first find the initial moment when \textit{velocityTI} exceeds a threshold of $0.05 m/s$ and stays above it for at least 3 consecutive frames.
    \item \textbf{Determine the pinch start time:} From the identified onset, we search backward to locate the point where \textit{velocityRot} has been continuously below $0.1\ deg/s$ for at least 5 frames. This time index is then identified as the exact start of a pinch action.
\end{itemize}



\begin{figure*}
    \centering
    \includegraphics[width=1.0\linewidth]{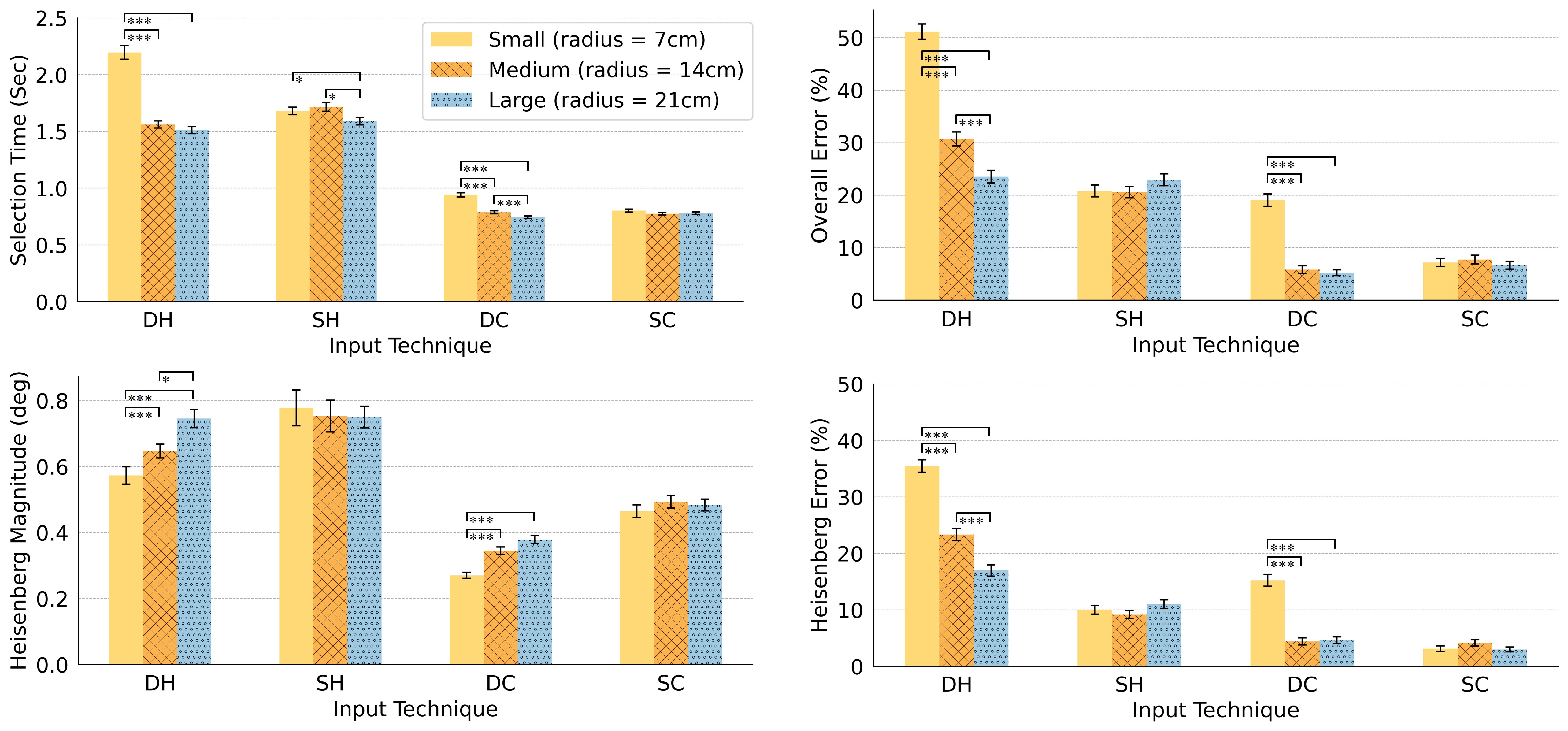}
    \caption{The performance summary of each technology under different target width conditions, including selection time, overall error, Heisenberg Error and Heisenberg Magnitude. The error bar represents the standard error. DH: direct selection with hand input, SH: score-based selection with hand input, DC: direct selection with controller input, SC: score-based selection with controller input.}
    \label{fig:quanresult1}
\end{figure*}

\subsection{Procedure}
We adopted a within-subjects design with a 
4 × 3 × 3 factorial structure: 4 \textbf{TECHNIQUE} conditions, 3 \textbf{TARGET WIDTH} levels, and 3 \textbf{TARGET SPACING} levels, following the grid layout from Lu et al.~\cite{BubbleRay:2020:VR}. The order of TECHNIQUE conditions was counterbalanced with a Latin square to reduce order effects. The combinations of TARGET WIDTH and TARGET SPACING were randomized across participants. Each condition was repeated three times to mitigate potential learning effects. In each scene associated with a specific TECHNIQUE condition, 49 (7 × 7 grid) spherical targets were displayed in a grid layout at a fixed distance of 8 meters from the participant, which follows prior work~\cite{HeisenbergEffect:CHI:2020} to observe the Heisenberg Effect clearly. The spheres were colored either red or blue. Also, following the experiment environment in BubbleRay~\cite{BubbleRay:2020:VR}, only the 25 inner targets (5 $\times$ 5 grid, excluding boundary targets) were used for selection tasks to ensure each target had four adjacent neighbors and an equal selection difficulty.

For each TECHNIQUE condition in a single trial, participants were instructed to sequentially select 6 red-colored targets, which were randomly chosen from the 5 $\times$ 5 inner grid. Before the formal experiment, participants received instructions and were given sufficient time (approximately 2 minutes per technique) to practice. They could verbally indicate when they felt ready or ask questions if they encountered any confusion. During the formal sessions, rest periods (at least 1 minute, or longer until participants reported no fatigue) were provided between trials. All participants completed the experiment while seated in a chair with armrests for arm support. The entire session lasted approximately 40 to 50 minutes per participant, including a semi-structured post-experiment interview where participants reflected on their experiences with the different selection techniques across conditions.

%% file: 4-Result.tex
\section{Result}
In total, 2592 trials (648 trials per input technique), including 21621 selection data points (7275 in DH, 5347 in SH, 4612 in DC, 4387 in SC), were collected. We initially removed 32 trials (1.24\% of 2592,  12 in DH, 11 in SH, 3 in DC and 6 in SC) where the completion time exceeded three standard deviations from the mean for each interaction technique × target width × target spacing condition. These outliers were all due to some participants
requesting short breaks and the oral reports accidentally triggering
a selection event upon entering a new scene. The Shapiro-Wilk test indicated that the collected data regarding selection time were not normally distributed, leading us to conduct a repeated-measures ANOVA using Aligned Rank Transform~\cite{ART:2021:UIST}. Post-hoc pairwise comparisons were performed using paired t-tests with Bonferroni correction. The primary performance metric under different target widths is shown in Figure~\ref{fig:quanresult1}, while the comparison under different target spacing is shown in Figure~\ref{fig:quanresult2}. We denote statistical significance with * for $p<0.05$, ** for $p<0.01$, and *** for $p<0.001$. A post-hoc sensitivity analysis ($\alpha = 0.05$, $power = 0.80$, $N = 24$, within-subject design) indicated that our design was powered to detect \textbf{medium effect sizes} (Cohen's $f \ge 0.26$, equivalent to $\eta^2 \ge 0.06$). Notably, the observed effect sizes span a wide range from small ($\eta^2 = 0.01$) to large ($\eta^2 = 0.35$) over the whole study. Statistically significant results characterized by small effect sizes should be interpreted with caution.

\subsection{Quantitative Results}

\label{QuanResult}
\subsubsection{Selection Time}

The effect of \textbf{TARGET WIDTH} on selection time is shown in Figure~\ref{fig:quanresult1}. \textbf{For DH}, an RM-ANOVA revealed a significant effect of target width ($F_{2,609.06} = 87.856$, $p < 0.001$, $\eta^2 = 0.22$), with the small targets taking significantly longer than both medium ($t_{609} = 10.719$, $p < 0.001$) and large targets ($t_{609} = 12.108$, $p < 0.001$). \textbf{For SH}, target width had a significant effect ($F_{2, 610.8} = 5.366$, $p < 0.01$, $\eta^2 = 0.02$), where the large targets yielded shorter selection times than medium ($t_{610} = 2.892$, $p < 0.05$) and small targets ($t_{610} = 2.782$, $p < 0.05$). The difference between small and medium targets ($t_{610} = 0.107$, $p = 0.994$) was not significant. \textbf{For DC}, a significant effect was observed ($F_{2, 618.01} = 84.866$, $p < 0.001$, $\eta^2 = 0.22$), with small targets producing longer selection time than both medium ($t_{618} = 8.976$, $p < 0.001$) and large targets ($t_{618} = 12.632$, $p < 0.001$). Medium poses longer selection time than large targets ($t_{618} = 3.636$, $p < 0.001$). \textbf{For SC}, the effect of target width was not significant ($F_{2,615.02} = 2.2264$, $p = 0.109$), indicating stable selection performance across all target widths.

\begin{figure*}[h]
    \includegraphics[width=1\linewidth]{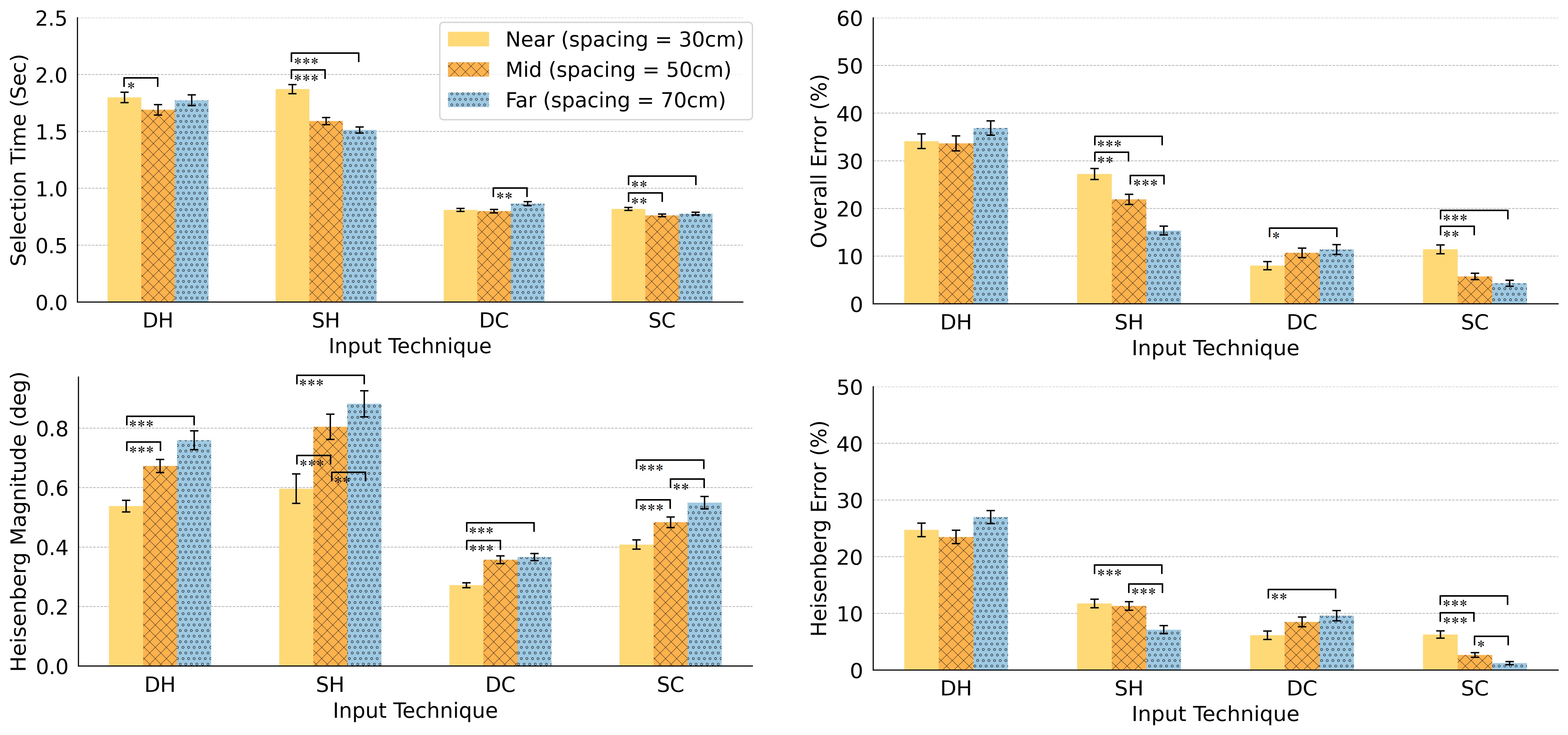}
    \caption{The performance summary of each technology under different target spacing conditions, including selection time, overall error, Heisenberg Error and Heisenberg Magnitude. The error bar represents the standard error. DH: direct selection with hand input, SH: score-based selection with hand input, DC: direct selection with controller input, SC: score-based selection with controller input.}
    \label{fig:quanresult2}
\end{figure*}

The effect of \textbf{TARGET SPACING} on selection time is shown in Figure~\ref{fig:quanresult2}. \textbf{For DH}, an RM-ANOVA revealed a significant effect of target spacing ($F_{2,609.08} = 3.371$, $p < 0.05,$, $\eta^2 = 0.01$). Only the spacing in near yielded longer selection time than mid ($t_{609} = 2.596$, $p < 0.05$). There is no significant effect between spacing in near and far ($t_{609} = 1.247$, $p = 0.426$), while spacing in mid and far ($t_{609} = -1.345$, $p = 0.371$). \textbf{For SH}, target spacing had a significant effect ($F_{2,610.1} = 45.874$, $p < 0.001$, $\eta^2 = 0.13$), the spacing in near yielded longer selection time than the spacing in mid ($t_{610} = 7.134$, $p < 0.001$) and far ($t_{610} = 9.105$, $p < 0.001$). \textbf{For DC}, an RM-ANOVA revealed a significant effect of target spacing ($F_{2,618.02} = 5.021$, $p < 0.01$, $\eta^2 = 0.02$). Spacing in mid yielded shorter selection time than the spacing in far ($t_{618} = -3.122$, $p < 0.01$). There is no significant effect between spacing in near and mid ($t_{618} = 1.088$, $p = 0.522$), while spacing in near and far ($t_{618} = -2.035$, $p = 0.105$). \textbf{For SC}, target spacing had a significant effect ($F_{2,615.02} = 10.063$, $p < 0.001$, $\eta^2 = 0.03$), where the spacing in near yielded longer selection time than the spacing in mid ($t_{609} = 4.356$, $p < 0.01$) and far ($t_{609} = 3.098$, $p < 0.01$).

\begin{table*}[!thb]
\centering
\caption{The proportion of the Heisenberg Error in the overall error across spatial variations. DC and DH consistently exhibit the high proportions, while score-based methods show lower and more stable proportions across target width and spacing.}
\label{tab:heisenberg_proportion}
\begin{tabular}{l|ccc|ccc|c}
\toprule
\multicolumn{1}{c}{} & \multicolumn{7}{c}{The proportion of the Heisenberg Error in Overall Error(\%)} \\
\midrule
\multicolumn{1}{c}{} & \multicolumn{3}{c}{ Target Width} & \multicolumn{3}{c}{Target Spacing} \\
\cmidrule(lr){2-8}
Input & Small (14cm) & Medium (28cm) & Large (42cm) & Near (30cm) & Mid (50cm) & Far (70cm) & Overall \\
\midrule
DC & 79.84 & 75.51 & 88.87 & 76.78 & 79.44 & 84.14 & 80.57 \\
SC & 43.48 & 53.51 & 44.89 & 54.82 & 46.18 & 28.31 & 47.50\\
DH & 69.30 & 75.92 & 72.17 & 72.47 & 69.76 & 73.18 & 71.86\\
SH & 48.10 & 44.43 & 47.94 & 43.14 & 51.58 & 46.35 & 46.86\\
\bottomrule
\end{tabular}
\end{table*}

\subsubsection{Overall Error}

The effect of \textbf{TARGET WIDTH} on Overall Error is shown in Figure~\ref{fig:quanresult1}. \textbf{For DH}, an RM-ANOVA revealed a significant effect of target width ($F_{2,609.06} = 166.4$, $p < 0.001$, $\eta^2 = 0.35$), with the small targets posing significantly higher error rate than both medium ($t_{609} = 12.678$, $p < 0.001$) and large targets ($t_{609} = 17.695$, $p < 0.001$), also the medium targets posing higher error than large targets ($t_{609} = 4.998$, $p < 0.001$). \textbf{For SH}, target width revealed no significant effect ($F_{2,610.17} = 1.334$, $p = 0.2642$). \textbf{For DC}, a significant effect was observed ($F_{2,618.04} = 93.736$, $p < 0.001$, $\eta^2 = 0.23$), with small targets producing higher selection error than both medium ($t_{618} = 11.847$, $p < 0.001$) and large targets ($t_{618} = 11.863$, $p < 0.001$). There is no significant effect between medium and large targets ($t_{618} = -0.058$, $p = 0.998$). \textbf{For SC}, the effect of target width was not significant ($F_{2,615.04} = 0.424$, $p = 0.655$). 


The effect of \textbf{TARGET SPACING} on Overall Error is shown in Figure~\ref{fig:quanresult2}. \textbf{For DH}, an RM-ANOVA revealed no significant effect of target spacing  ($F_{2,609.09} = 1.201$, $p = 0.302$). \textbf{For SH}, target spacing had a significant effect ($F_{2,610.24} = 36.02$, $p < 0.001$, $\eta^2 = 0.11$), where the spacing in near posed higher selection error than spacing in mid ($t_{610} = 3.557$, $p < 0.01$) and far ($t_{610} = 4.901$, $p < 0.001$). The spacing in mid produced a higher selection error than far($t_{610} = 8.452$, $p < 0.001$). The spacing in mid posed a higher selection error than far ($t_{610} = 4.901$, $p < 0.001$). \textbf{For DC}, an RM-ANOVA revealed a significant effect of target spacing ($F_{2,618.05} = 3.854$, $p < 0.05$, $\eta^2 = 0.01$). The spacing in near posed a lower error than far ($t_{618} = -2.578$, $p < 0.05$). There is no significant effect between spacing in near and mid targets ($t_{618} = 2.180$, $p = 0.075$), while spacing in mid and far ($t_{618} = 0.401$, $p = 0.915$). \textbf{For SC}, target spacing had a significant effect ($F_{2,615.04} = 30.389$, $p < 0.001$, $\eta^2 = 0.09$), where the spacing in near posed a higher selection error than spacing in mid ($t_{615} = 5.407$, $p < 0.001$) and far ($t_{615} = 7.562$, $p < 0.001$). No significant effect between spacing in mid and far ($t_{615} = -2.134$, $p = 0.084$).

\subsubsection{Heisenberg Magnitude}

The effect of \textbf{TARGET WIDTH} on Heisenberg Magnitude is shown in Figure~\ref{fig:quanresult1}. 
\textbf{For DH}, an RM-ANOVA revealed a significant effect ($F_{2,609.07} = 24.241$, $p < 0.001$, $\eta^2 = 0.07$), with the small targets produced lower Heisenberg Magnitude than both medium ($t_{609} = -4.053$, $p < 0.001$) and large targets ($t_{609} = -6.929$, $p < 0.001$). The medium objects produced lower Heisenberg Magnitude than the large objects ($t_{609} = -2.868$, $p < 0.05$). 
\textbf{For SH}, target width had no significant effect ($F_{2,610.09} = 0.184$, $p = 0.832$).
\textbf{For DC}, a significant effect was observed ($F_{2,618.03} = 43.178$, $p < 0.001$, $\eta^2 = 0.12$), with small targets posing higher Heisenberg Error than both medium ($t_{618} = -6.830$, $p < 0.001$) and large targets ($t_{618} = -8.870$, $p < 0.001$). No significant difference is found between medium and large ($t_{618} = -2.067$, $p = 0.098$).
\textbf{For SC}, the effect of target width was not significant ($F_{2,615.01} = 1.686$, $p = 0.186$).

The effect of \textbf{TARGET SPACING} on Heisenberg Magnitude is shown in Figure~\ref{fig:quanresult2}. \textbf{For DH}, an RM-ANOVA revealed a significant effect ($F_{2,609.07} = 26.328$, $p < 0.001$, $\eta^2 = 0.08$), with the spacing in near produced lower Heisenberg Magnitude than both spacing in mid ($t_{609} = -5.330$, $p < 0.001$) and spacing in far ($t_{609} = -6.928$, $p < 0.001$). No significant difference is found between spacing in mid and spacing in far ($t_{609} = -1.604$, $p = 0.2445$).
\textbf{For SH}, target spacing had a significant effect ($F_{2,610.1} = 82.482$, $p < 0.001$, $\eta^2 = 0.21$), with the spacing in near produced lower Heisenberg Magnitude than both spacing in mid ($t_{610} = -9.251$, $p < 0.001$) and spacing in far ($t_{610} = -12.344$, $p < 0.001$). A significant difference is also found between spacing in mid and spacing in far ($t_{610} = -3.102$, $p < 0.01$).
\textbf{For DC}, target spacing had a significant effect ($F_{2,618.03} = 34.877$, $p < 0.001$, $\eta^2 = 0.10$), with the spacing in near produced lower Heisenberg Magnitude than both spacing in mid ($t_{618} = -6.517$, $p < 0.001$) and spacing in far ($t_{618} = -7.780$, $p < 0.001$). No significant difference is found between spacing in mid and far ($t_{618} = -1.271$, $p = 0.412$).
\textbf{For SC}, target spacing had a significant effect ($F_{2,615.01} = 38.674$, $p < 0.001$, $\eta^2 = 0.11$), with the spacing in near produced lower Heisenberg Magnitude than both spacing in mid ($t_{615} = -4.982$, $p < 0.001$) and far ($t_{615} = -8.766$, $p < 0.001$). The spacing in mid also produced a lower Heisenberg Magnitude than far ($t_{615} = -3.758$, $p < 0.01$).

\subsubsection{Heisenberg Error}

The effect of \textbf{TARGET WIDTH} on Heisenberg Error is shown in Figure~\ref{fig:quanresult1}. \textbf{For DH}, an RM-ANOVA revealed a significant effect ($F_{2,609.12} = 92.737$, $p < 0.001$, $\eta^2 = 0.23$), with the small targets posing higher Heisenberg Error than both medium ($t_{609} = 8.428$, $p < 0.001$) and large targets ($t_{609} = 13.477$, $p < 0.001$), also the medium targets posing higher Heisenberg Error than large targets ($t_{609} = 5.033$, $p < 0.001$). \textbf{For SH}, target width reveals no significant effect ($F_{2,610.33} = 1.477$, $p = 0.229$). \textbf{For DC}, a significant target width effect was observed ($F_{2,618.05} = 71.457$, $p < 0.001$, $\eta^2 = 0.19$), with small targets producing higher Heisenberg Error than both medium ($t_{618} = 10.744$, $p < 0.001$) and large targets ($t_{618} = 9.907$, $p < 0.001$). No significant effect is found between medium and large ($t_{618} = -0.800$, $p = 0.703$). \textbf{For SC}, the effect of target width was not significant ($F_{2,615.09} = 1.885$, $p = 0.15$).

The effect of \textbf{TARGET SPACING} on Heisenberg Error is shown in Figure~\ref{fig:quanresult2}. \textbf{For DH}, an RM-ANOVA revealed no significant effect ($F_{2,609.16} = 2.594$, $p = 0.0756$). \textbf{For SH}, target spacing had a significant effect ($F_{2,610.47} = 14.362$, $p < 0.001$, $\eta^2 = 0.04$), where the spacing in near yielded a higher Heisenberg Error than spacing in far ($t_{610} = 4.520$, $p < 0.001$). No significant effect is found between spacing in near and spacing in mid ($t_{610} = 0.237$, $p = 0.969$). The spacing in mid also posed a higher Heisenberg Error than the spacing in far ($t_{610} = 4.754$, $p < 0.001$). \textbf{For DC}, an RM-ANOVA revealed a significant effect of target spacing with a small effect size ($F_{2,618.06} = 5.170$, $p < 0.01$, $\eta^2 = 0.02$). The spacing in the near produced a lower Heisenberg Error than the spacing in far ($t_{618} = -3.148$, $p < 0.01$). No significant effect is found between spacing in near and spacing in mid ($t_{618} = -2.140$, $p = 0.083$), while spacing in mid and spacing in far ($t_{618} = -1.011$, $p = 0.570$). \textbf{For SC}, target spacing had a significant effect ($F_{2,615.08} = 30.534$, $p < 0.001$, $\eta^2 = 0.09$), where the spacing in near yielded a higher Heisenberg Error than spacing in mid ($t_{615} = 5.013$, $p < 0.001$) and spacing in far ($t_{615} = 7.672$, $p < 0.001$). The spacing in mid posed a higher Heisenberg Error than spacing in far ($t_{615} = 2.660$, $p < 0.05$).

We also illustrate the percentage of the Heisenberg Error in the Overall Error under different spatial layouts in Table~\ref{tab:heisenberg_proportion}.

\begin{figure*}
    \includegraphics[width=1\linewidth]{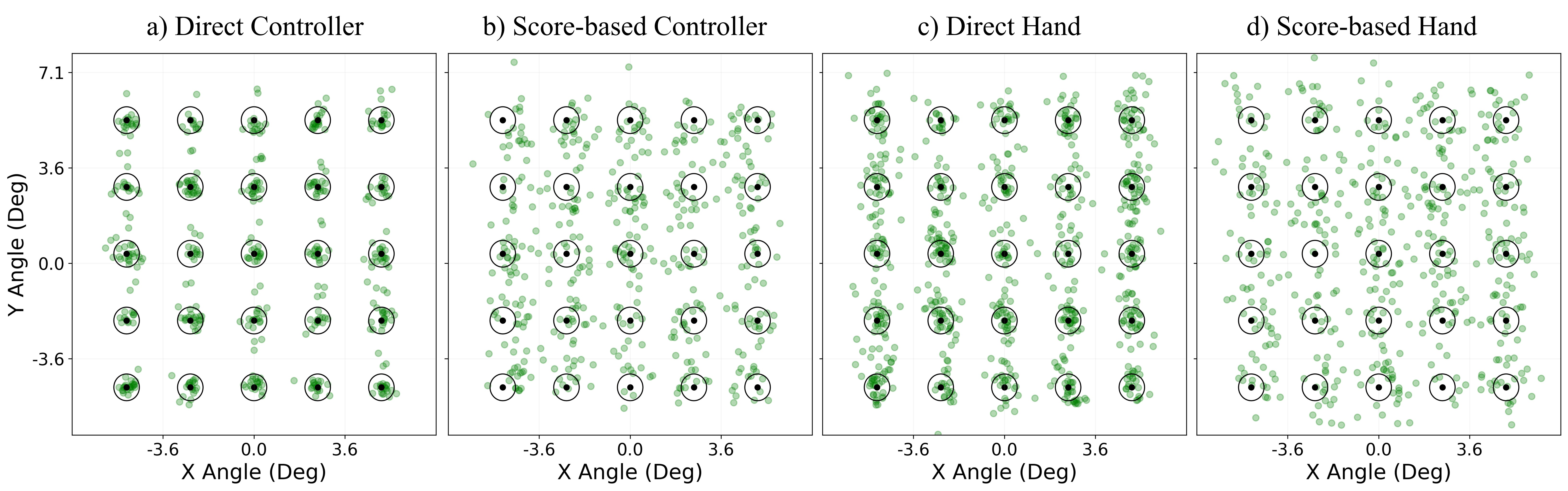}
    \caption{Distribution of selection endpoints across input techniques with a target width of 28 cm and spacing of 70 cm. Endpoints under direct selection (a, c) are more concentrated around the targets, whereas score-based selection (b, d) shows greater dispersion. These endpoints represent the raw pointing ray rather than the final snap-to raycasting.}
    \label{fig:endpoint_distribution}
\end{figure*}

\subsubsection{Directional Bias with Heisenberg Effect}

To further quantify the systematic directional offset introduced by the Heisenberg Effect, we analyzed the spatial displacement for each selection. This offset was calculated as the angular distance between the aiming vector at the initiation of a discrete action (i.e., press or pinch) and the vector at its completion. Each selection's final position was categorized relative to the initial aiming vector into one of four quadrants: TopRight, TopLeft, BottomRight, or BottomLeft. The aggregated distribution of these directional offsets is presented in Table~\ref{tab:directiondistribution}.

\begin{table}[htbp]
\centering
\caption{Direction distribution of Heisenberg Magnitude. Most input shifts are concentrated in the \textbf{TopRight}, especially for SC and DC, and the \textbf{BottomLeft} exhibits the lowest frequency across all inputs.}
\label{tab:directiondistribution} 
\begin{tabular}{lcccc}
\toprule
& \multicolumn{4}{c}{Shift Direction (\%)} \\
\cmidrule(lr){2-5}
Input &TopRight&TopLeft&BottomRight&BottomLeft \\
\midrule
DC & 39.89 & 28.03 & 16.83 & 15.11\\
SC & 47.41 & 29.89 & 10.75 & 11.31\\
DH & 29.68 & 31.60 & 25.04 & 13.69\\
SH & 34.90 & 33.31 & 21.64 & 10.14 \\
\bottomrule
\end{tabular}
\end{table}

Consistent with the findings of Wolf et al~\cite{HeisenbergEffect:CHI:2020}, our results reveal a distinct directional bias for controller-based interactions. We observed a consistent upward shift in the final selection point for 72.6\% of trials using the Quest 3 controller. We attribute this systematic error to the physical location and actuation mechanics of the trigger button, which tends to cause a slight upward rotation of the controller during a press. In contrast, the spatial offset for bare-hand input showed no dominant directional bias and was significantly more distributed. We hypothesize that this variance stems from the inherent diversity in individual hand physiology and the unique pinch gestures adopted by each participant, leading to less predictable displacement vectors upon selection.


\subsection{Evaluation and Discussion}

\subsubsection{General Characteristics of the Heisenberg Effect}


The quantitative analysis of Heisenberg Magnitude shows that for direct selection, the Heisenberg Magnitude decreases as the target width decreases. Prior work attributed this phenomenon to anticipatory postural adjustments that modulate muscle tension and arm posture depending on the perceived target width~\cite{PosAdjust:2008:EBR,HeisenbergEffect:CHI:2020}. Our study extends this understanding that, regardless of whether participants performed direct or score-based selection, the Heisenberg Magnitude also decreases as target spacing becomes smaller. This suggests that when task difficulty increases, participants reinforce motor control strategies to deal with higher selection error rates, thereby reducing the impact of the Heisenberg Effect. This interpretation is supported by post-hoc interviews, in which the majority of participants (22/24) reported elevated nervousness when performing the most challenging tasks.

\subsubsection{Bare-Hand vs. Controller Input} 


Consistent with prior research~\cite{HeisenbergEffect:CHI:2020}, direct controller input is subject to the Heisenberg Effect. Our findings indicate that the performance limitations of this effect are most pronounced only for very small targets, as we found no significant difference in the Overall Error difference when selecting medium versus large objects.

In quantitative evaluation, we found DH produces longer selection time ($t_{1254} = 49.383$, $p < 0.001$), higher overall error ($t_{1254} = 26.200$, $p < 0.001$), Heisenberg error ($t_{1254} = 21.910$, $p < 0.001$) and Heisenberg magnitude ($t_{1254} = 26.169$, $p < 0.001$) than DC. As well, SH produces longer selection time ($t_{1252} = 65.570$, $p < 0.001$), higher overall error ($t_{1252} = 20.725$, $p < 0.001$), Heisenberg error ($t_{1252} = 13.523$, $p < 0.001$) and Heisenberg magnitude ($t_{1252} = 14.324$, $p < 0.001$) than SC. The \textbf{H1} is supported by the analysis of two different input modalities, which reveals that bare-hand input is more susceptible to the Heisenberg Effect than controller input. While the absolute number of Heisenberg Errors was greater for bare-hand input, they constituted a smaller proportion of total errors compared to the controller condition. This suggests that, beyond the Heisenberg Effect, the initial pointing accuracy of bare-hand input before the selection gesture is lower than that of controller-based inputs.

We found that in the period before pinch action, slight head shaking or camera movement may also disturb hand tracking. Unlike controller inputs, which leverage IMU-based tracking for relatively more stable spatial localization, hand tracking relies on vision-based approaches and is sensitive to camera motion. Consequently, minor head movements can result in unexpected spatial shifts of hand input. Nevertheless, following the definition of the Heisenberg Effect established in prior work~\cite{HeisenbergEffect:CHI:2020}, we did not categorize it as part of the Heisenberg Error.

\subsubsection{Direct vs. Score-based Selection}



In quantitative evaluation, we found DH produces higher overall error ($t_{1246} = 12.946$, $p < 0.001$), Heisenberg error ($t_{1246} = 19.273$, $p < 0.001$) and lower Heisenberg magnitude ($t_{1246} = 3.056$, $p < 0.05$) than SH. As well, DC produces longer selection time  ($t_{1260} = 3.165$, $p < 0.01$), higher overall error ($t_{1260} = 3.888$, $p < 0.001$), Heisenberg error ($t_{1260} = 8.034$, $p < 0.001$) and lower Heisenberg magnitude  ($t_{1260} = 14.470$, $p < 0.001$) than SC. The \textbf{H1} is further supported by the analysis of two different selection mechanisms. Overall, score-based selection offers a noticeable performance improvement over direct selection, especially in the case of hand selection to reduce the selection error. 

Additionally, the quantitative analysis of performance metrics in Section~\ref{QuanResult} also supports \textbf{H2}. Target width significantly affected performance and Heisenberg-related metrics for direct selection, with medium significant effects found. It can be concluded that direct selection is more sensitive to target width, but its performance only drops drastically when the targets are small. In contrast, target spacing had a pronounced impact on score-based selection, particularly in dense environments, leading to increased overall error and Heisenberg Error. These results indicate that spatial factors influence the Heisenberg Effect in distinct ways depending on the selection mechanism. Different from direct selection, score-based selection is largely unaffected by target width due to its avoidance of null selections. However, densely packed targets significantly impair its performance, leading to higher selection errors. Interestingly, score-based selection produces greater Heisenberg Magnitude than direct selection.

\subsubsection{Visual Indicator Matters in the Heisenberg Effect}

When some participants (7/24) felt that score-based selection was inferior to direct selection, they noted it to “\textit{the sticky ray cannot determine the endpoint exactly}.” Our observations support the counterintuitive viewpoint: during score-based selection, the scores of nearby objects often approach that of the intended target, making it difficult to distinguish. Although the curvature of the Bezier curve can partly indicate the original direction of ray and its alignment with the target, participants still struggled to control the ray to prevent unintended shifts.

Further evidence of the role of visual indicators in target selection is shown in Figure~\ref{fig:endpoint_distribution}, showing the distribution of endpoints for different input techniques. Endpoints of direct selection cluster tightly around targets, whereas score-based selection produces a more scattered endpoint distribution. This contrast suggests that visual indicators are critical for stabilizing endpoint control. In direct selection, the clear spatial alignment between the ray and the target provides strong visual cues, enabling participants to converge on their intended objects. In score-based selection, the absence of explicit spatial anchoring leads to higher endpoint variability, especially in high-density environments. These findings underscore that the presence and clarity of visual indicators influence both user performance and confidence in target selection.

\subsubsection{Bias Caused by Different Devices and Setups}

Compared to quantitative results reported by Wolf et al.~\cite{HeisenbergEffect:CHI:2020}, our measured Heisenberg Magnitude for direct controller tracking (avg = 0.332°) is slightly lower than theirs (avg $>$ 0.5°). We hypothesize that this discrepancy stems from hardware differences: our study employed Meta Quest 3, a newer VR device with different button placements and pressure thresholds, whereas Wolf et al. used the HTC Vive. Additional testing is needed to evaluate how button position and device design influence directional shifts across different VR devices with controllers.

Our results indicate that the Heisenberg Effect accounts for 80.57\% of overall errors in target selection tasks, markedly higher than the 30.49\% reported in ballistic selection tasks~\cite{HeisenbergEffect:CHI:2020}, and closely aligned with errors observed in stationary selection tasks (81.98\%). This discrepancy is likely due to our use of visual highlighting around the intended object, which reduced initial positioning errors~\cite{RayCursor:2019:CHI}. It suggests that the Heisenberg Effect is also shaped by the visual highlighting of intended targets and the feedback provided during selection. 

Although Target Width and Target Spacing were treated as main effects in our study, post-hoc analysis revealed a nuanced interaction effect for Overall Error. With the large width (42 cm), the overall error is significantly higher at spacing in near than spacing in far ($t_{2524} = 2.74$, $p < 0.05$). Notably, no significant interaction effect was observed in selection time and Heisenberg-related metrics.

%% file: 5-MitigationTechniques.tex
\section{Proposed Mitigation Method}
\begin{figure*}[h]
    \centering
    \includegraphics[width=1\linewidth]{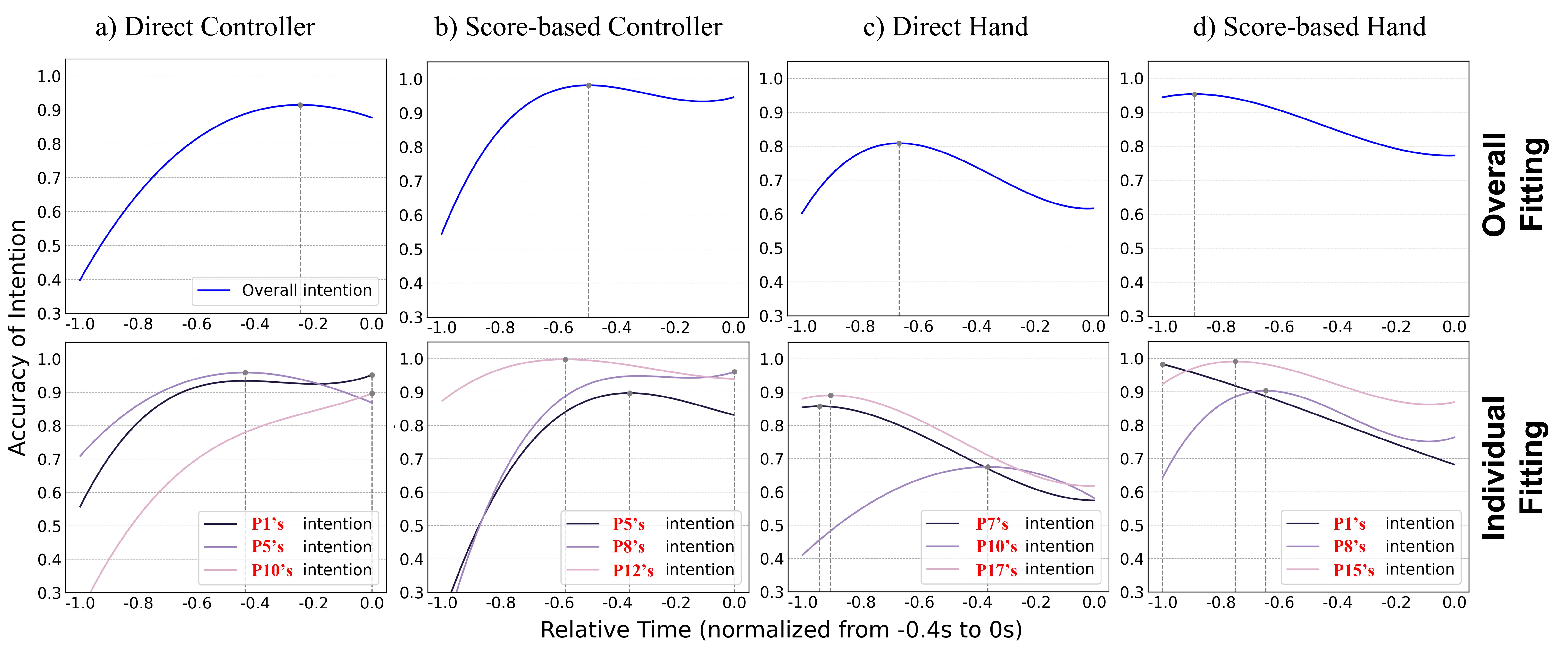}
    \caption{Trends of user intention accuracy over time across four selection modalities and mechanisms. Each curve is fitted with a third-degree polynomial within 0.4s before selection. Relative time (x axis) is normalized across selection events, where \textbf{0.0 denotes selection completion}. Accuracy of Intention (y axis) is defined as the probability that the user's selection ray is pointing at or snapping to the correct target object, where \textbf{1.0 denotes that the intended object must be the target one.} The first row shows overall accuracy curves trained on data from P1–P19, illustrating general trends. The bottom row highlights individual participants whose accuracy patterns deviate from the overall trend.}
    \label{fig:relativeTime}
\end{figure*}

\subsection{Observation and Motivation on Intention History}
\label{temporalanalysis}
To analyze how user intention evolves prior to selection, we normalize the temporal dimension of each selection event. Interaction frames stored in the history cache are rescaled to a 0–1 range based on their relative positions within the event timeline. This normalization accounts for instances where the duration of a selection event lasted shorter than the 0.4-second threshold. We then split the collected data into a training set (19 participants, 17187 selection records) and a test set (5 participants, 4157 selection records). Observing that intention accuracy in the preceding time window follows an S-shaped trend, we model the likelihood of correctly matching the intended object to the target using a third-degree polynomial for all input methods. The formulas for normalization of time length and fitting function for intention accuracy are as follows:
\begin{equation}
\tilde{T} = \frac{t - T_c}{T_{e}},
\end{equation}
\begin{equation}
f(\tilde{T}) = a \tilde{T}^{3} + b \tilde{T}^{2} + c \tilde{T} + d,
\end{equation}
where $\tilde{T}$ represents the relative time of a candidate item within the intention history for each selection. $T_c$ denotes the timestamp at which the user performs the confirmation action, and $t$ is the timestamp of the candidate item being evaluated. $T_e=0.4s$ represents the expiration window that defines the valid temporal range for recording past interaction frames. The coefficients $a$, $b$, $c$, and $d$ are empirically fitted based on the training data, and the overall fitting function is shown in the first row of Figure~\ref{fig:relativeTime}, demonstrating the varied selection accuracy in the interaction process. 

Consistent with the multiple process model that includes searching, aiming, and confirmation~\cite{MPM:2025:CHI}, intention accuracy for all methods first increases, then drops sharply during the confirmation stage. As shown in Figure~\ref{fig:relativeTime} b) and d), it indicates a strong influence of the Heisenberg Effect on selection performance, which is particularly pronounced in the DH and SH conditions (nearly \textbf{20\% accuracy loss}). Notably, the score-based hand input achieved an accuracy of nearly 98\%, which is comparable to two controller conditions, indicating that incorporating historical information can significantly improve the accuracy by alleviating the Heisenberg effect. 

While the Heisenberg Effect introduces unavoidable perturbations during the final motor action, users’ pointing behavior before confirmation is often more stable and better reflects their true intention. This observation motivates a history-based mitigation strategy: rather than relying solely on the instantaneous pointing state at the final selection timestamp, we aggregate evidence from a short temporal cache window preceding confirmation. By weighting past pointing samples, the system can infer the most likely intended target even when the final action is transiently disrupted to a false selection.

However, a unified weighting function is insufficient. As shown in the second row of Figure~\ref{fig:relativeTime}, individual intention accuracy trends can deviate significantly from the overall model, even though most participants conform to the general pattern. This variability stems from both fine-grained factors (e.g., hand tremor, pinch speed) and coarse-grained factors (e.g., hand posture). Our findings suggest that common compensation mechanisms, which backtrack a selection to a fixed time frame as proposed in prior work~\cite{HeisenbergEffect:CHI:2020}, can be suboptimal. Such a direct backtracking method often increases false positive errors in direct selections (i.e., DC and DH). Our temporal analysis of intention accuracy indicates that historical data should be weighted differently to mitigate the Heisenberg Effect. 

\subsection{Weighted VOTE Based on Empirical Data}

To integrate our findings to improve the performance in target selection tasks, we firstly referred to previous history-based techniques~\cite{VOTE:2018:IJHCS, BackTracer:2023:IJHCS} and implemented the classic VOTE methods~\cite{VOTE:2018:IJHCS} as a baseline. VOTE achieves the final target selection by accumulating intention over a temporal cache window (0.4s in our implementation) prior to the confirmation event and selecting the object that receives the highest number of votes, with each frame pose one pointing sample within the window contributing equally. The classic VOTE formulation is as follows:
\begin{equation}
O^* = \arg\max_{o \in \tilde{O}} 
\sum_{\substack{\tilde{T} \leq 1}} 
\mathbf{1}[V(\tilde{T}) = o],
\end{equation}
where $O^*$ denotes the final object returned by Weighted VOTE. $\tilde{O}$ represents the finite set of all candidate (selectable) objects in the intention history. $V(\tilde{T})$ indicates the object indicated by the underlying interaction technique at time $\tilde{T}$, equals the special symbol ``null'' when no object is currently indicated. $1\mathbf[$·$]$ indicates a function returns 1 if the condition is true, 0 otherwise.

Our approach is inspired by the principle of intention accumulation, but extends it by moving beyond the equal-weight assumption. Instead of treating all historical frames equally, we weight them according to our empirical findings (Fig.~\ref{fig:relativeTime}). We introduce~\textbf{Weighted VOTE} tailored to mitigate disturbances caused by the Heisenberg Effect.
Considering our findings, \textbf{each time frame requires different
voting weights}: the higher the historical accuracy in a specific time frame, the higher the voting weights should be. In order to enhance the precisions of selection and give valuable frames higher voting weights,  we applied a sigmoid function to enhance the contrast of voting weights. The weighting mechanism can be formulated as: 
\begin{equation}
W(\tilde{T}) = \frac{1}{1 + e^{-k(f(\tilde{T}) - A)}},
\end{equation}
\begin{equation}
O^* = \arg\max_{o \in \tilde{O}} 
\sum_{\substack{\tilde{T} \leq 1}} 
W\left( \tilde{T} \right) \cdot \mathbf{1}[V(\tilde{T}) = o],
\end{equation}
where the $W(\tilde{T})$ indicates a temporal weighting function that assigns a non-negative weight to each vote based on its relative time. $k$ indicates an adjustable parameter controlling the steepness of the sigmoid weighting curve, assigned with 10 in our implementation. $A$ denotes the shift parameter of the sigmoid. For each fitted weight function, it is adaptively defined as the empirical two-thirds quantile of the function values within the $[0, 1]$ interval.

To address individual differences, we developed the Adaptive Weighted VOTE for a more personalized backtrack strategy. Specifically, we fitted a cubic function using the first 20\% of the test user’s selection history and interpolated it with the overall fitted function (P1–P19), assigning a higher interpolation weight (i.e., 0.4 in our implementation) to the individual model. This approach allows the system to adapt more rapidly to personal behavior patterns while retaining the stability of the general model.

\subsection{Evaluation on Mitigation Strategies}

In addition to the original method, VOTE, and our proposed Weighted VOTE, we selected the temporal point corresponding to the peak of the fitted function on the training data as the backtracking reference time frame. We refer to this approach as ``Shift to Before''. After fitting the voting weights for each input method, we performed a comparative evaluation on both the original selection and the improved techniques. To ensure fair evaluation across all mitigation strategies, we use the last 80\% of the data in the test set for evaluation. 

\begin{table}[htbp]
\centering
\caption{Evaluation of selection error with mitigation strategies in the test set. The bold indicates the technique with the lowest error rate. Our proposed techniques, including the Weighted VOTE and Adaptive Weighted VOTE, achieve a lower error rate compared to the other baselines.}
\label{tab:mitigationeval}
\begin{tabular}{lccccc}
\toprule
& \multicolumn{5}{c}{Error Rate in the Test Set (\%)} \\
\cmidrule(lr){2-6}
Input & Origin & \makecell{Shift to \\ Before} & VOTE & \makecell{Weighted \\ VOTE} & \makecell{Adaptive \\ Weighted \\ VOTE}\\
\midrule
DC & 15.20 & 12.06 & 3.49 & 2.96 & \textbf{2.94}\\
SC & 13.35 & 14.72 & 11.56 & \textbf{8.18} & 8.56\\
DH & 32.92 & 25.00 & 10.42 & \textbf{10.42} & 10.65 \\
SH & 21.96 & 8.22 & 11.90 & 7.93 & \textbf{7.54}\\
\bottomrule
\end{tabular}
\end{table}

As shown in Table~\ref{tab:mitigationeval}, Weighted VOTE significantly reduces the error rate across all input techniques, though its impact varies. The most substantial improvement was observed in SH, where applying Weighted VOTE reduced its error rate to a level comparable with controller tracking. This success stems from the longer aiming times inherent to bare-hand interaction, which provides a richer history of user intention that VOTE leverages effectively. Conversely, Weighted VOTE provided minimal benefits for direct selection techniques (DH and DC). For these methods, historical data was already dominated by the correct target and numerous empty objects. Since VOTE is designed to ignore empty objects by default, its application offered no significant improvements. Such features of voting-based methods also explain why Adaptive Weighted VOTE did not demonstrate outstanding performance in direct selections. Only a minor performance improvement was found in the SH technique. We discuss adaptive techniques in Section~\ref{dis_contexthistory}.


%% file: 6-Discussion.tex
\section{Discussion and Future Work}

In this study, we conduct a systematic analysis of how the Heisenberg Effect matters with various input techniques and introduce a potential mitigation strategy based on intention data. This section demonstrates findings and implications for future work. 

\subsection{The Manifestations and Mitigation Principles of the Heisenberg Effect in Spatial Computing}
\label{dis_heisenberg}
Prior analysis of the Heisenberg Effect has examined direct ray input with handheld controllers, identifying target width as the dominant factor~\cite{HeisenbergEffect:CHI:2020}. Our results extend this account across both input modalities and selection mechanisms, showing that the manifestation of the Heisenberg Effect is strongly technique-dependent.

For direct selection, performance is dominated by the target width for both controller and hand input. This pattern matches classical motor control models of pointing, where movement behavior is governed by the speed–accuracy tradeoff in Fitts' Law~\cite{FittsLaw:1954:JEP}. As target width decreases, increased precision demands lead users to rely on corrective movements, which are susceptible to signal-dependent noise. As a result, the observed pointing behavior is increasingly shaped by the selection task itself rather than the user’s intention, yielding a width-driven Heisenberg Effect. Score-based selection, by contrast, exhibited a distinct sensitivity toward target density rather than width. It reflects a fundamental change in task structure: spatial precision is partially offloaded to scoring functions, reducing endpoint accuracy requirements but transforming selection into a decision process under uncertainty. Similar effects have been reported in probabilistic and techniques, where geometric constraints are relaxed at the cost of increased competition among nearby candidates~\cite{BubbleCursor:CHI:2005,MMSelection:IJHCS:2009}. From a motor control perspective, score-based selections do not eliminate the speed–accuracy tradeoff but redistribute its causation. Optimal and feedback-based control models describe motor behavior as a continuous balance between predictive planning and noisy execution, with variability emerging during both planning and correction~\cite{HMP:PR:1988,OFC:NatureN:2002}. Under dense layouts, planning uncertainty and motor noise create minor directional disturbances that can significantly alter relative target rankings. This leads to selection instability, producing a density-driven Heisenberg Effect.

Prior history-based methods implicitly assume all intention records before the final selection contribute equally~\cite{BackTracer:2023:IJHCS,VOTE:2018:IJHCS}. Still, our analysis reveals that the Heisenberg Effect introduces an inherent asymmetry of intention in the motor control process. Earlier ray orientations often provide more reliable evidence of user intent than the final endpoint. Weighted VOTE acts as a temporal inference mechanism rather than a direct heuristic. It integrates directional evidence over time while discounting frames that are likely to introduce motor execution noise. By further adjusting weighting profiles to observed behavioral patterns, Adaptive Weighted VOTE accommodates individual variability. More fundamentally, our strategies can be interpreted as a technique-agnostic temporal inference module rather than a standalone input technique. Weighted VOTE is compatible with various selection techniques, including ray-based pointing~\cite{HeisenbergEffect:CHI:2020}, hand-based~\cite{RestfulRaycast:2025:DIS}, gaze-based selection~\cite{GazePinch:2017:SUI}, and score-based methods~\cite{BubbleRay:2020:VR,FocalPoint:IMWUT:2023,FocalSelect:2024:TVCG}. 


\subsection{Limitation and Future Work}
\label{dis_contexthistory}

Our work focuses on the analysis of the Heisenberg Effect, and we have found that it appears differently in the interaction technique and underlying motor or decision processes. Future work could further expand the analysis of the Heisenberg Effect to other input modalities, including compound inputs like Hand + Eye~\cite{GazePinch:2017:SUI} and alternative methods such as head-ray or foot-based selection~\cite{GazePinch:2017:SUI,FootTextEntry:2024:CHI}. The effect should also be evaluated across different hardware to ensure consistent interaction quality~\cite{Heisenbergv1:2001:HCII,DualStick:2025:VR}. By further decomposing motor control theory, future research could explore visual feedback designs. Alternative indicators or decoupling visual feedback from selection mechanisms may mitigate performance degradation in dense environments~\cite{FocalSelect:2024:TVCG}. For the mitigation strategy, our study did not examine the user experience with subjective data (e.g., NASA-TLX, SUS, or perceived confidence). Future work may improve adaptive variants of Weighted VOTE by incorporating additional behavioral inputs, such as kinematic data from eye tracking~\cite{RIDS:2022:UIST}. Moreover, integrating real-time error detection~\cite{ErrorDetection:2025:CHI} could enable automatic labeling of intention history and support dynamic, human-in-the-loop adaptation of the Weighted VOTE.

%% file: 7-Conclusion.tex
\section{Conclusion}
In this work, we evaluate the Heisenberg Effect in target selection tasks across four widely used VR interaction techniques, focusing on their impact on selection performance and user experience. To assess the influence of environmental factors, we conducted a user study (\textit{N} = 24) with nine combinations of target width $\times$ target spacing. Results indicate that direct ray-casting performs poorly with small targets, while the score-based technique is less effective in densely spaced environments. The Heisenberg Effect appears to be a primary factor underlying the inferior performance of bare-hand input compared to controllers. Building on insights from a temporal analysis of accuracy in historical intentions, we introduce a weighted VOTE method to enhance selection performance. Our evaluation demonstrates that it significantly improves the accuracy of score-based selections with bare-hand input, which is comparable to that of controller input.